\documentclass[letterpaper,twocolumn,10pt]{article}
\usepackage{usenix2019_v3}

\usepackage{tikz}
\usepackage{amsmath}

\usepackage{mathdots}
\usepackage{yhmath}
\usepackage{cancel}
\usepackage{color}
\usepackage{siunitx}
\usepackage{array}
\usepackage{multirow}
\usepackage{amssymb}
\usepackage{gensymb}
\usepackage{tabularx}
\usepackage{booktabs}
\usetikzlibrary{fadings}
\usetikzlibrary{patterns}
\usetikzlibrary{shadows.blur}

\usepackage{filecontents}
\usepackage{todo}
\usepackage{svg}
\usepackage{graphicx}
\begin{document}

\date{}

\title{\Large \bf QPEP: A QUIC-Based Approach to Encrypted Performance Enhancing Proxies for High-Latency Satellite Broadband}

\author{
{\rm James Pavur}\\
Oxford University
\and
{\rm Martin Strohmeier}\\
armasuisse
\and
{\rm Vincent Lenders}\\
armasuisse
\and
{\rm Ivan Martinovic}\\
Oxford University
} 
\maketitle

\begin{abstract}
	Satellite broadband services are critical infrastructures enabling advanced technologies to function in the most remote regions of the globe. However, status-quo services are often unencrypted by default and vulnerable to eavesdropping attacks. In this paper, we challenge the historical perception that over-the-air security must trade off with TCP performance in high-latency satellite networks due to the deep-packet inspection requirements of Performance Enhancing Proxies (PEPs).
	
	After considering why prior work in this area has failed to find wide adoption, we present an open-source encrypted-by-default PEP - QPEP - which seeks to address these issues. QPEP is built around the open QUIC standard and designed so individual customers may adopt it without ISP involvement. QPEP's performance is assessed through simulations in a replicable docker-based testbed. Across many benchmarks and network conditions, QPEP is found to avoid the perceived security-encryption trade-off in PEP design. Compared to unencrypted PEP implementations, QPEP reduces average page load times by more than 30\% while also offering over-the-air privacy. Compared to the traditional VPN encryption available to customers today, QPEP more than halves average page load times. Together, these experiments lead to the conclusion that QPEP represents a promising new approach to protecting modern satellite broadband connections.

\end{abstract}

\section{Introduction}
Historically, security and performance have often traded-off in satellite broadband service design. As a result, many satellite internet service providers (ISPs) do not offer over-the-air encryption in their satellite networks, exposing sensitive customer data to eavesdropping and manipulation. The physical characteristics of satellite networks have given rise to this trade off as techniques to optimize TCP connections over high-latency satellite links require deep-packet inspection which is incompatible with end-to-end encryption. This causes VPNs and traditional encrypted tunnels to perform poorly over satellite networks.

Since the early 2000s, academics and satellite communications business have struggled with the challenge of encrypting these TCP communications while maintaining acceptable performance. In Section~\ref{sec:background} we delve into this challenge and several notable approaches to better understand why proposed solutions have seen limited adoption in real-world networks. We find that, even when encrypted satellite broadband products are available, they are generally geared towards large business clients and inaccessible to individual home users of satellite broadband. Moreover, available offerings are often costly and unverifiable due to their proprietary nature, disincentivizing adoption. 

We find that existing academic proposals, while numerous, are often purely theoretical or lack replicable source-code. To our knowledge, no open-source encryption tool exists for performant TCP communications over satellite links. Even if it did, the lack of standardized testing and simulation environments for PEP products make it difficult to compare and benchmark techniques without direct access to a satellite ISP and their ground stations. Additionally, the limited number of open-source unencrypted PEPs available mean that potential researchers seeking to make a practical security contribution must either repurpose outdated code (generally written in C/C++) to implement modern encryption protocols or must re-invent PEP proxies from scratch, requiring deep knowledge not just of security but also low-level network programming.

The end result is that satellite broadband users have no good options. They (or their ISPs) must purchase expensive and unvetted proprietary applications, accept the substantial performance hit caused by general-purpose VPNs, or transmit sensitive data in clear text over massive satellite signal footprints.

This paper seeks to address both the lack of encryption options and the high barriers to research in this domain. Its primary contribution is QPEP - an open-source PEP which offers privacy by default. Built around the open and modern QUIC transportation protocol, QPEP benefits from research innovations in a much larger community than the niche satellite security field. The tool itself is implemented entirely in Go, an accessible and modern language, to encourage research contributions from developers who may be uncomfortable implementing encryption and networking protocols in C/C++. As a secondary contribution, the paper presents an open-source and extensible docker-based simulation testbed, built on the OpenSAND satellite networking simulation engine~\cite{dockerDocker2020,viveristechnologiesOpenSAND2019}. The testbed is provided to facilitate easy replication of our own performance benchmarks and assess the performance of other PEP appliances. It enables meaningful comparisons between PEP approaches and aims to decrease the friction involved in future contributions which improve on our methods.

Within this testbed, we find that QPEP performs well across several benchmarking tests. It more than halves average page load times compared to the use of VPN encryption and matches or exceeds the performance of unencrypted PEP applications. Additional simulations are conducted under adverse network conditions and orbital configurations and we assess the viability of some simple modifications to the standard QUIC protocol for satellite network usage. Ultimately, we find that QPEP represents a promising solution for over-the-air encryption in satellite networks which can be deployed by individual satellite customers without sacrificing network performance or security.

\section{Background and Related Work}
\label{sec:background}
To understand the security and performance trade-off in satellite broadband services requires a basic understanding of the design of satellite networks and the behavior of the TCP protocol in this environment. In this section, we provide a high-level overview of the key factors behind this dynamic and prior work in academia and industry to address them.

\subsection{Satellite Networks and Eavesdropping}
\label{sec:satellite_networking}
\begin{figure}
	\includegraphics[width=\linewidth]{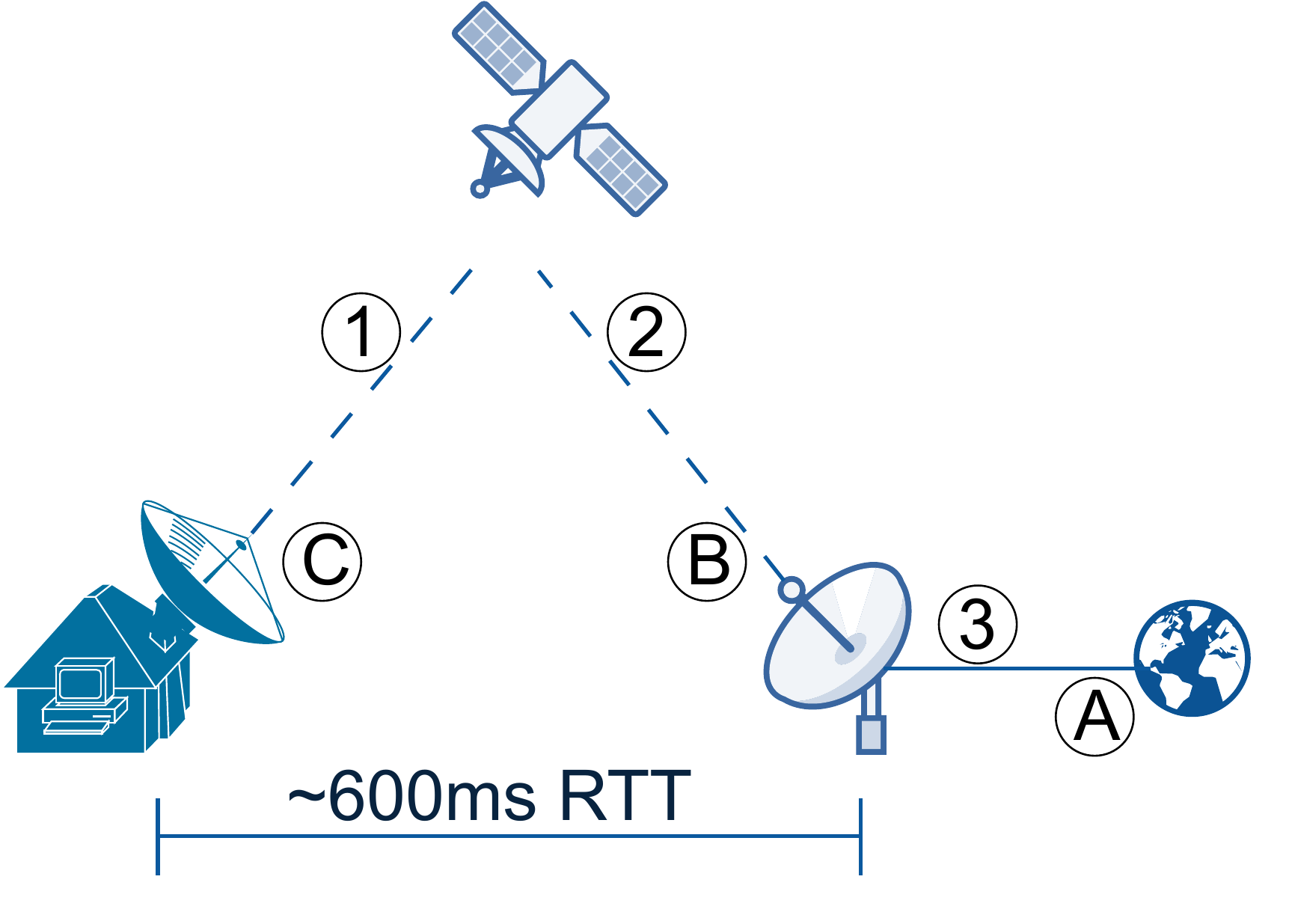}
	\caption{Notional Overview of a GEO Network}
	\label{fig:bent_pipe}
\end{figure}
Our principal focus is on satellite broadband provided from platforms in geostationary earth orbit (GEO). The basic operation of GEO broadband can be thought of as a ``bent pipe'' (see Figure~\ref{fig:bent_pipe}). First, the broadband customer points their satellite at a fixed position in the sky and transmits IP packets towards a receiving satellite transponder. These IP packets are encapsulated using a protocol like Return Link Encapsulation (RLE) over Digital Video Broadcasting - Return Channel Via Satellite (DVB-RCS) or S-band Mobile Interactive Multimedia (S-MIM)~(step 1 in Figure~\ref{fig:bent_pipe}). The satellite then relays this encapsulated stream towards a ground station on Earth (step 2). At the ground station, the satellite ISP will decapsulate the stream into IP packets and route them across the internet via terrestrial networks (step 3). When a response directed to the satellite customer is received by the ISP (step A in Figure~\ref{fig:bent_pipe}), this process occurs again in reverse. The ISP encapsulates the stream using one of a number of forward-link standards (e.g. Generic Stream Encapsulation (GSE) over Digital Video Broadcasting - Satellite (DVB-S)) and transmits the stream up towards the satellite (step B). The satellite relays this stream down to the customer whose modem decapsulates it and routes it along their home or business network  (step C).

As GEO is located more than 30,000~km away from the Earth's surface, a single satellite has line of sight to a vast area on the surface (theoretically as much as 40\% of the Earth's surface, but practically closer to 20\% for broadband communications). This has the advantage of making GEO broadband a relatively inexpensive mechanism of providing global service. Only a half-dozen satellites are needed for almost complete Earth coverage (barring some polar areas). 

This vast footprint also creates intuitive appeal for eavesdropping attacks as wide-reaching signals are not targeted towards specific user terminals. As a result, an attacker monitoring GEO broadcasts can passively eavesdrop on traffic belonging to an entire continent of satellite customers. Real world experiments have shown that it is possible to use inexpensive equipment to intercept deeply sensitive data from these feeds - ranging from login credentials for offshore wind farms to passport information belonging to cargo ship crews~\cite{laurieAtelliteHackingFun2009,egeaPlayingSatelliteEnvironment2010,pavurSecretsSkyPrivacy2019a,pavurTaleSeaSky2020}.

In light of such security concerns, it is not intuitively clear why status quo satellite broadband services fail to encrypt customer traffic over the air. At its core, the barrier to encryption is a physical one. Speed of light delays over the 30,000~km hop to GEO are substantial and round-trip packet latency can exceed 600~ms in an otherwise perfectly optimized network. While many of these challenges could theoretically be mitigated by using satellites closer to Earth in low Earth orbit (LEO), this increases the complexity of global coverage. LEO satellites benefit from only 50~ms in speed-of-light latency but only maintain line of sight to a point on Earth's surface for a matter of minutes. To provide consistent global coverage with 50~ms latency would require hundreds or even thousands of satellites. In the status quo, even LEO constellations can experience round-trip packet delays of up to 1,500~ms depending on the specific route a message takes through the constellation to a ground station~\cite{mcmahonMeasuringLatencyIridium2005}. Thus, while many new LEO constellations are in the early stages of development, and have made ambitious claims regarding expected latency, GEO broadband and high-latency will both likely remain relevant characteristics for both modern and upcoming satellite networks~\cite{spacexStarlinkMission2020,onewebOneWeb}.

\subsection{TCP Performance Over Satellite}
\label{sec:tcp_satellite_performance}
To understand how this high latency prevents robust encryption in satellite networks, one must consider the impact of latency on TCP performance. Without any optimizations, TCP connections over satellite links would appear sluggish and unusable for most customers - even with high bandwidth allocation and top-quality equipment due to these physical effects. In this subsection, we will touch on two of the main contributors to this performance degradation. However, these barriers and many others have also been much more extensively characterized in prior work~\cite{oueslati-boulahiaTCPSatelliteLinks2000,cainiTCPPEPDTN2009,bisioPerformanceEnhancedProxy2009}.

\subsubsection{Barriers to TCP in Satellite Networks}
The first challenge to TCP performance in satellite networks arises from the requirement that TCP data packets are responded to with an acknowledgment (ACK) message. These ACK exchanges cause additional blocking round-trip transmissions over the high-latency satellite link. The effect is compounded by the three-way handshake used for session initialization which, in the best case, takes upwards of 1,500~ms to complete over GEO. When visiting a website with embedded images and related files, many three-way handshakes be required to download the relevant content - compounding delays. Although modern TCP implementations may employ ACK decimation to reduce the total number of ACKs, these are not configured to appropriate ratios for satellite networks. Indeed, in many older and embedded devices, ACKs are still elicited by every TCP packet, massively increasing the perceived latency of the satellite link.

The second challenge arises from the nature of TCP congestion control and the TCP ``slow-start'' session initialization. In order to implement effective ACK decimation, TCP slow-start gradually increases the ratio of data segments to ACKs until a desired congestion window is reached. The time this process takes to finish is thus a function of round-trip times (RTT) over the satellite link. Even once a connection has reached optimal window size, packet loss will be misidentified as a sign of network congestion and cause the slow-start sequence to restart from the beginning. While modern satellite transmissions are more reliable than in the past, packet loss is still common compared to terrestrial networks. As a result, TCP sessions are both slow to maximize their bandwidth usage, and, once maximized, struggle to maintain that state.

These are two factors among many impacting TCP performance over satellite links. Beyond latency effects, particularities in satellite network topology combine to create a uniquely hostile environment for standard TCP implementations. As a result, satellite service providers have had to build and deploy modifications to the TCP traffic on their networks. It is these modifications that act as substantial barriers to the effective deployment of over-the-air encryption services.

\subsubsection{PEPs}
The most common approach to optimizing TCP traffic over satellite environments is the use of a class of appliances called ``Performance Enhancing Proxies'' or PEPs and loosely described in IETF RFC 3135. While PEPs can differ substantially and many implementations are proprietary, most PEPs adhere to a few basic principles.

There are two typical PEP deployment options - integrated or distributed. In integrated PEPs, a PEP appliance operates on a single endpoint - typically the ISP satellite gateway between the satellite network and the internet. In distributed PEPs, a PEP appliance operates on multiple endpoints - typically both on the customer satellite modem and the ISP gateway.

In either deployment, the PEP observes TCP traffic which passes through it and applies optimizations in order to compensate for satellite performance issues. Typically, PEPs do this in a manner which is invisible to the endpoints of the conversations so that no modifications are required on consumer hardware and the ISP can use commercial internet routing technology. This is referred to as a ``transparent'' PEP. However, the concept of transparency here is somewhat misleading as, in many cases, PEP modifications are still detectable (e.g. changing TCP sequence numbers).

Beyond this, PEPs vary quite broadly. The modifications made to optimize TCP traffic are often proprietary and implementation-specific. One common approach is to ``split'' incoming TCP connections prior to transmission across the satellite link and issue local ACK messages immediately for received TCP packets. This allows three-way handshakes and congestion control to be negotiated locally before the satellite hop but requires the PEP developer to implement logic ensuring that packet loss and connection errors on one side of the split are correctly handled on the other.

In distributed PEPs, this splitting approach is extended to create a tunnel between the individual PEP installations (see Figure~\ref{fig:pep_overview}). A TCP packet arriving at the client-side PEP (e.g. on the home satellite modem), is terminated locally as a TCP connection, and the payload is then forwarded through GEO using a modified TCP protocol (e.g. with larger congestion windows) or an alternative non-TCP protocol optimized for satellite networks. At the ISP gateway, a second PEP receives this modified TCP packet, converts it back to normal traffic, and sends it along a locally-managed TCP connection to the internet. This allows for ACK messages to be handled local to either end of the satellite hop and reduces the number of round-trips needed to initialize a TCP session.

\tikzset{every picture/.style={line width=0.75pt}} 
\begin{figure}
	\centering

\tikzset{every picture/.style={line width=0.75pt}} 

\begin{tikzpicture}[x=0.75pt,y=0.75pt,yscale=-1,xscale=1]

\draw    (20,9.67) -- (19.5,289.57) ;

\draw    (100,10.1) -- (99.5,290) ;

\draw    (220,9.67) -- (219.5,289.57) ;

\draw    (300,10.1) -- (299.5,290) ;

\draw    (19.5,18.43) -- (96.65,50.66) ;
\draw [shift={(98.5,51.43)}, rotate = 202.67000000000002] [color={rgb, 255:red, 0; green, 0; blue, 0 }  ][line width=0.75]    (10.93,-3.29) .. controls (6.95,-1.4) and (3.31,-0.3) .. (0,0) .. controls (3.31,0.3) and (6.95,1.4) .. (10.93,3.29)   ;

\draw    (98.5,51.43) -- (21.38,79.74) ;
\draw [shift={(19.5,80.43)}, rotate = 339.84000000000003] [color={rgb, 255:red, 0; green, 0; blue, 0 }  ][line width=0.75]    (10.93,-3.29) .. controls (6.95,-1.4) and (3.31,-0.3) .. (0,0) .. controls (3.31,0.3) and (6.95,1.4) .. (10.93,3.29)   ;

\draw    (19.5,80.43) -- (96.63,109.72) ;
\draw [shift={(98.5,110.43)}, rotate = 200.79] [color={rgb, 255:red, 0; green, 0; blue, 0 }  ][line width=0.75]    (10.93,-3.29) .. controls (6.95,-1.4) and (3.31,-0.3) .. (0,0) .. controls (3.31,0.3) and (6.95,1.4) .. (10.93,3.29)   ;

\draw  [dash pattern={on 4.5pt off 4.5pt}]  (98.5,51.43) -- (218.55,79.97) ;
\draw [shift={(220.5,80.43)}, rotate = 193.37] [color={rgb, 255:red, 0; green, 0; blue, 0 }  ][line width=0.75]    (10.93,-3.29) .. controls (6.95,-1.4) and (3.31,-0.3) .. (0,0) .. controls (3.31,0.3) and (6.95,1.4) .. (10.93,3.29)   ;

\draw    (220.5,80.43) -- (297.65,112.66) ;
\draw [shift={(299.5,113.43)}, rotate = 202.67000000000002] [color={rgb, 255:red, 0; green, 0; blue, 0 }  ][line width=0.75]    (10.93,-3.29) .. controls (6.95,-1.4) and (3.31,-0.3) .. (0,0) .. controls (3.31,0.3) and (6.95,1.4) .. (10.93,3.29)   ;

\draw    (299.5,113.43) -- (222.38,141.74) ;
\draw [shift={(220.5,142.43)}, rotate = 339.84000000000003] [color={rgb, 255:red, 0; green, 0; blue, 0 }  ][line width=0.75]    (10.93,-3.29) .. controls (6.95,-1.4) and (3.31,-0.3) .. (0,0) .. controls (3.31,0.3) and (6.95,1.4) .. (10.93,3.29)   ;

\draw    (220.5,142.43) -- (297.65,174.66) ;
\draw [shift={(299.5,175.43)}, rotate = 202.67000000000002] [color={rgb, 255:red, 0; green, 0; blue, 0 }  ][line width=0.75]    (10.93,-3.29) .. controls (6.95,-1.4) and (3.31,-0.3) .. (0,0) .. controls (3.31,0.3) and (6.95,1.4) .. (10.93,3.29)   ;

\draw    (298.5,175.43) -- (220.42,198.85) ;
\draw [shift={(218.5,199.43)}, rotate = 343.3] [color={rgb, 255:red, 0; green, 0; blue, 0 }  ][line width=0.75]    (10.93,-3.29) .. controls (6.95,-1.4) and (3.31,-0.3) .. (0,0) .. controls (3.31,0.3) and (6.95,1.4) .. (10.93,3.29)   ;

\draw  [dash pattern={on 4.5pt off 4.5pt}]  (218.5,199.43) -- (100.44,229.93) ;
\draw [shift={(98.5,230.43)}, rotate = 345.52] [color={rgb, 255:red, 0; green, 0; blue, 0 }  ][line width=0.75]    (10.93,-3.29) .. controls (6.95,-1.4) and (3.31,-0.3) .. (0,0) .. controls (3.31,0.3) and (6.95,1.4) .. (10.93,3.29)   ;

\draw    (99.5,230.43) -- (21.42,253.85) ;
\draw [shift={(19.5,254.43)}, rotate = 343.3] [color={rgb, 255:red, 0; green, 0; blue, 0 }  ][line width=0.75]    (10.93,-3.29) .. controls (6.95,-1.4) and (3.31,-0.3) .. (0,0) .. controls (3.31,0.3) and (6.95,1.4) .. (10.93,3.29)   ;

\draw (66,28) node  [font=\small,rotate=-21.8] [align=left] {SYN};
\draw (162,54) node  [font=\small,rotate=-13.51] [align=left] {SYN (Copied)};
\draw (260,87) node  [font=\small,rotate=-21.33] [align=left] {SYN (Copied)};
\draw (65,88) node  [font=\small,rotate=-21.8] [align=left] {ACK};
\draw (265,152) node  [font=\small,rotate=-21.8] [align=left] {ACK};
\draw (54,59) node  [font=\small,rotate=-338.92] [align=left] {SYN-ACK};
\draw (256,122) node  [font=\small,rotate=-338.92] [align=left] {SYN-ACK};
\draw (252,181) node  [font=\small,rotate=-344.35] [align=left] {DATA};
\draw (159,205) node  [font=\small,rotate=-344.35] [align=left] {DATA};
\draw (58,276) node  [font=\footnotesize] [align=left] {\textit{Workstation}\\\textit{to Sat-Modem}};
\draw (260,274) node  [font=\footnotesize] [align=left] {\textit{Groundstation}\\\textit{to Internet}};
\draw (158,277) node  [font=\footnotesize] [align=left] {\textit{High-Latency}\\\textit{Satellite Link}};
\draw (53,236) node  [font=\small,rotate=-344.35] [align=left] {DATA};

\end{tikzpicture}

\caption{Split Distributed PEP Handshake Example}
\label{fig:pep_overview}
\end{figure}
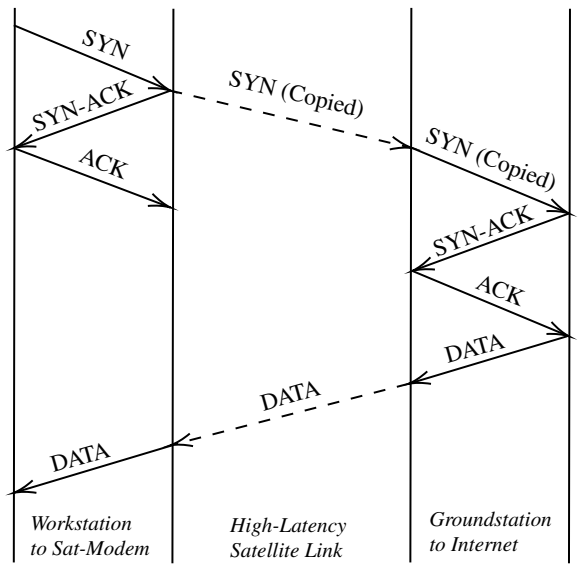

Other PEP optimizations can range from modifying TCP congestion control to bundling related packets into single transmissions. Commercial implementations often offer higher-level features such as inspecting HTTP payloads and combining requests for web-pages with their associated content. A substantial body of existing work on PEPs covers these optimizations in detail not only for satellites, but also other latency sensitive environments (e.g. cellular networks)~\cite{oueslati-boulahiaTCPSatelliteLinks2000,cainiTCPPEPDTN2009,bisioPerformanceEnhancedProxy2009}.

\subsubsection{Security Consequences}
PEPs have become a vital component of satellite broadband and customers have come to expect the performance characteristics of PEP-accelerated networks. This has created unintended tension between broadband performance and security.

As noted in RFC 3135, PEPs break the end-to-end semantics of IP connections. Specifically, they require that the PEP appliance transparently modify packets ``over-the-wire'' - essentially acting as a benevolent man-in-the-middle on all TCP connections. This makes it impossible for PEPs to function on network layer encryption tunnels, such as IPSec VPNs, which encrypt the TCP header information necessary to split sessions and optimize packets. While VPNs still work over PEPs, performance for users is as if no PEP was deployed, creating functionally unusable services. As a result, end consumers are faced with a choice between the security and privacy of VPNs and the convenience and performance of PEPs.

In short, latency necessitates the use of PEPs and PEPs prevent the use of typical end-to-end encryption. The result is that many modern satellite networks still transmit deeply sensitive information over-the-air in clear text. These issues are, of course, fairly well known and many academic and industry techniques have been presented to mitigate them.

\subsection{Existing Security Approaches}
In this section, we will briefly consider some of the more consequential approaches to satellite broadband encryption at each layer of the TCP/IP protocol stack. This analysis better characterizes how, despite a long history of research, PEP-compatible security remains unsolved in practice.

\subsubsection{Physical and Link-Layer Approaches}
Many techniques for over-the-air encryption focus on the lower layers of the networking stack - before TCP/IP becomes relevant. For example, physical-layer techniques such as frequency hopping with channel patterns derived from cryptographic keys or direct sequence spread spectrum (DSSS) have been suggested as a mechanism for securing the entire satellite link from eavesdropping attacks~\cite{liSecureSpectrumefficientFrequency2016}. Likewise the injection of artificial noise into satellite signals as an alternative to key-based encryption has also been proposed~\cite{zhengPhysicalLayerSecurity2012}. These schemes tend to prioritize military communications systems. This makes sense as they often require expensive changes to groundstation and terminal hardware that would be unpalatable for commercial adopters.

At the link layer, proposals can still incur hardware costs but these costs are often restricted to hardware chips used for protocol decapsulation. For example, the Consultative Committee for Space Data Systems (CCSDS) has proposed Space Data Link Security (SDLS), a communications protocol with built-in encryption for telemetry commands to scientific space missions~\cite{ccsdsSpaceDataLink2018}. Likewise, in the satellite television broadcasting space, the proprietary Common Scrambling Algorithm (CSA) has long been used to restrict broadcast access to paying satellite television subscribers using smart-cards, albeit with notable security weaknesses~\cite{liSecurityAnalysisDVB2007}.

One security issue for link and physical layer encryption that is unique to broadband relates to the multi-user environment. It is rarely economically feasible to allocate each satellite terminal a unique channel or transponder. In practice, modems determine which packets in a given stream are relevant to their owners on the basis of packet header information (e.g. IP addresses). In such systems, customers who share a forward-link frequency must necessarily also share physical/link-layer keys needed to receive that frequency.

\subsubsection{Network and Transport-Layer Approaches}
\label{sec:pep_encrypt}
In order to provide over-the-air encryption with scalable per-customer keys, a number of network layer techniques have been proposed. In contrast with lower level approaches, interactions with TCP PEPs must now be considered directly. Many seek to replicate traditional VPN software with bespoke modifications - for example by creating a modified IPSec with special encapsulating headers visible to PEP appliances~\cite{duquerroySatiPSecOptimizedSolution2004, djeddaiIPSecOPEPIPSecPEPs2016}. Proprietary satellite VPNs also exist which, while public information on their design is limited, are likely similar in principle~\cite{encorenetworksBANDIT, dsdtelecomDSDSatelliteVPN}. Beyond concerns arising from the use of proprietary encryption schemes, these non-standard encapsulation layers can increase operator costs by limiting compatibility with the ISP's existing networking equipment.

As one of the principal challenges for satellite broadband encryption is PEP-compatibility, it makes intuitive sense to incorporate encryption within PEP appliances themselves - straddling the network and transport layers. This may be achieved by, for example, implementing an encrypted protocol over the satellite hop in a distributed PEP system. The transmitting PEP would first modify the TCP packets, then encrypt them. The receiving PEP would subsequently decrypt the received packets and forward them along the internet as normal. Many real-world PEP encryption products appear to employ this approach~\cite{newtecEL810MobilePEPBox2008,rignetCyphreLinkEndtoEndData, hughesHN7000SModemProduct}. 

To the best of our knowledge, most, if not all, implementations of encrypted PEPs are proprietary and security claims are thus difficult to independently verify. However, purported leaked manufacturer documents allude to built-in law-enforcement/intelligence back-doors in prominent encrypted PEP products~\cite{tellitecgmbh.TCSpyDocumentation2006}. Our own superficial analysis of one satellite router found numerous cryptographic shortcomings, such as Diffie-Hellman implementations which are susceptible to man-in-the-middle attacks and key/IV reuse that permits replay attacks. Similar vulnerabilities have been alluded to in prior research conducted in the mid-2000s~\cite{adelsbachInsiderAttacksEnabling2007}. Generally, the costs of adopting an encrypted PEP must be undertaken by the satellite ISP. ISPs may not perceive these purchases as value-for-money absent customer or regulatory pressure.

\subsubsection{Application-Layer Approaches}
An alternative approach would be the use of encryption protocols which operate over the PEP-accelerated TCP connection (i.e. encrypting payloads but leaving TCP headers exposed). The widespread use of TLS encryption for websites, for example, has the effect of encrypting customer data which is transmitted over-the-air on satellite links. However, this still leaks potentially sensitive data (such as the IPs a customer visits). Moreover, real-world observations of modern satellite traffic (see~\ref{sec:satellite_networking}) have found that, while customers could use TLS and other encrypted protocols, many still use insecure alternatives such as POP3 or HTTP.

While the decision to use insecure protocols is the customer's, satellite ISPs may nevertheless have a duty of care to protect this data over-the-air. One option would be to tunnel these insecure protocols into an application-layer encrypted stream before the data egresses from the client. This method is employed in some commercial satellite encryption products but has the notable disadvantage of requiring a software application to be installed on the client's computer~\cite{groundcontrolIGVPNUsingApplicaiton2003}. This limits compatibility with many embedded devices and creates friction in the customer experience.

\section{QPEP Design and Implementation}
\begin{figure*}
	\centering
	\includegraphics[width=\textwidth]{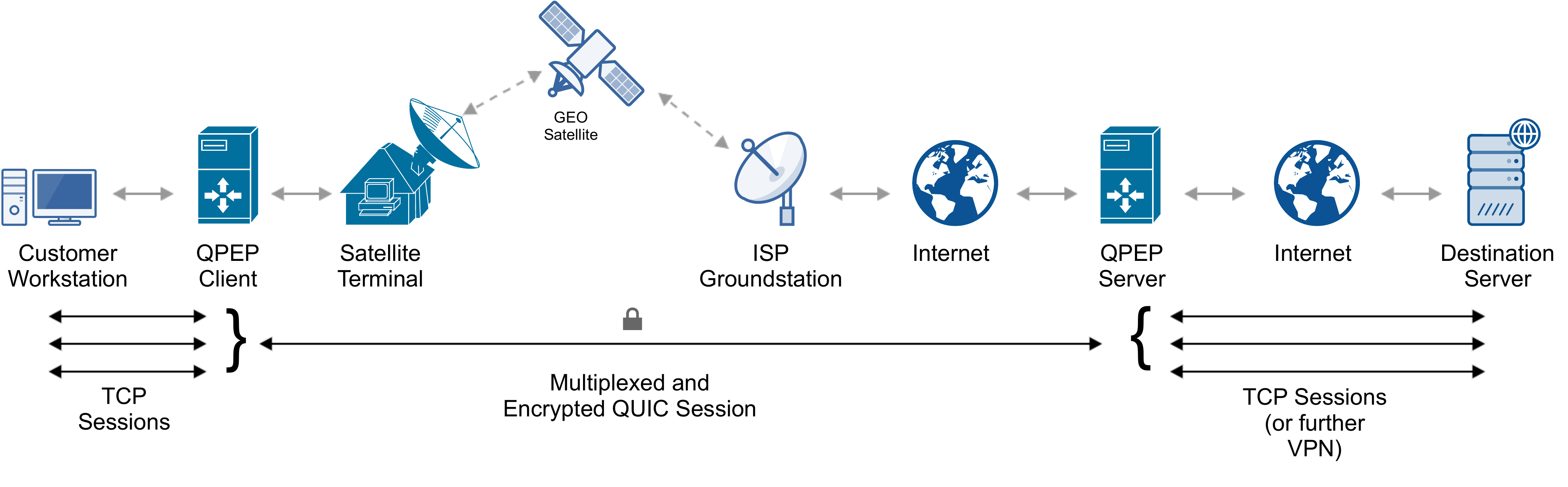}
	\caption{Simplified Overview of QPEP Architecture}
	\label{fig:qpep_overview}
\end{figure*}

To address some of these shortcomings in prior work and provide an open-source and non-proprietary tool for satellite encryption which can be used by individual satellite customers, we have developed QPEP. At its core, QPEP follows a distributed ``snooping'' PEP model similar to the methods described in Section~\ref{sec:pep_encrypt}. The QPEP client tunnels TCP traffic over the satellite link inside a stream that leverages the encrypted QUIC transport protocol. Tunneled traffic is decapsulated by a receiving QPEP server which then routes the decapsulated traffic over the internet as if it were the client. A high level overview of this architecture appears in Figure~\ref{fig:qpep_overview}. This section covers further details of the QPEP system and design decisions made in our proof-of-concept implementation.

\subsection{Use of the QUIC Protocol}

The principal difference between QPEP and existing proprietary implementations is the use of the QUIC protocol for tunneling traffic over the satellite link. While the QUIC protocol is still in the early stages of development and standards are evolving, it has seen real-world adoption in terrestrial networks due to its performance and security advantages over TCP. Several of these benefits make QUIC a promising option for deployment in the secure PEP context.

\subsubsection{QUIC Security Benefits}
QUIC is an encrypted-by-default protocol. Unlike TCP, the session initialization process for QUIC also incorporates a modified version of the TLS 1.3 handshake (see Figure~\ref{fig:quic_vs_vpn}). This means that every QUIC session provides both encryption for encapsulated payloads and built-in end-point authentication. The former mitigates against the security and privacy concerns raised by status-quo satellite broadband while the latter prevents the aforementioned man-in-the-middle issues in some proprietary encrypted PEPs.

A key factor distinguishing the use of a QUIC tunnel from commercial tunneling-based PEPs (e.g. Tellitec's ETCP product) is that QUIC, while theoretically a transport-layer protocol, is implemented in practice as an application layer protocol on top of IP and UDP~\cite{newtecEL810MobilePEPBox2008}. As a result, the use of QPEP does not require the satellite ISP to install a PEP server on their own gateway to decapsulate encrypted traffic. Instead, a satellite customer can use QUIC to tunnel their traffic through the ISP gateway and onwards to a QPEP server running on any internet-routable endpoint.

This means that individual satellite customers can employ QPEP without participation from their ISP. In this way, QPEP might best be compared to prior work on satellite VPN protocols~\cite{duquerroySatiPSecOptimizedSolution2004,djeddaiIPSecOPEPIPSecPEPs2016}. However, unlike these proposals, which often reveal limited portions of the TCP header to ISP PEPs (e.g. destination IP, port numbers, and TCP flags), only the IP address and UDP port of the customer's upstream QPEP server are visible. All information regarding the actual TCP connection is encapsulated within the QUIC tunnel. This allows customers to benefit from both satellite encryption and PEP acceleration without requiring any special trust in their ISP. In most satellite networks, QPEP should function even in the presence of other ISP-installed PEP appliances. Of course, customers may not be interested in configuring QPEP themselves. In such cases QPEP can also be installed by the satellite ISP onto the customer modem and their own gateway just as with traditional PEP appliances.

\subsubsection{QUIC Performance Benefits}
Beyond these security benefits - which would largely apply to any encrypted UDP tunnel - the use of QUIC offers several notable performance advantages.

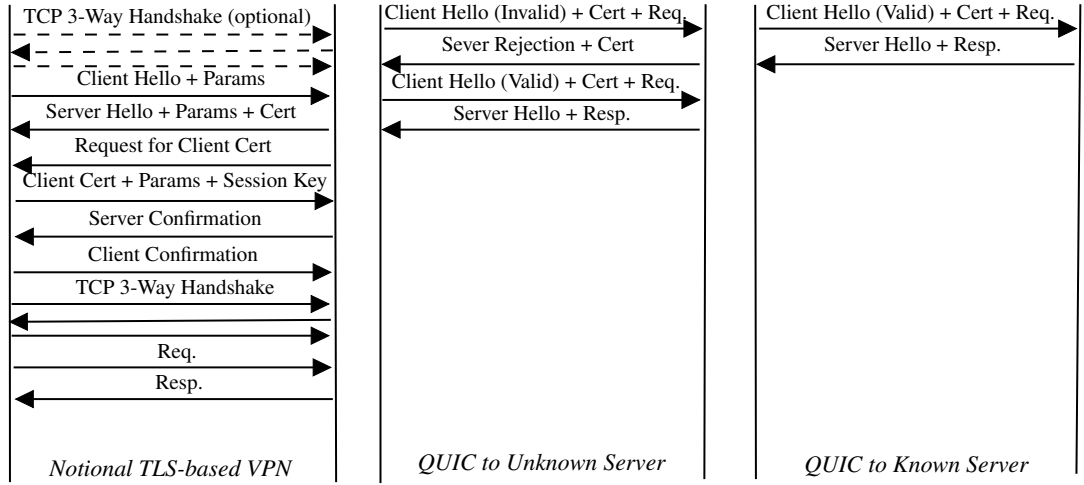
\begin{figure*}[h!]
	\centering

\tikzset{every picture/.style={line width=0.75pt}} 

\begin{tikzpicture}[x=0.75pt,y=0.75pt,yscale=-1,xscale=1]

\draw    (17.5,19.29) -- (17.65,260.2) ;

\draw    (181.5,19) -- (182.15,260.2) ;

\draw    (18.5,65.29) -- (175.5,65.29) ;
\draw [shift={(178.5,65.29)}, rotate = 180] [fill={rgb, 255:red, 0; green, 0; blue, 0 }  ][line width=0.08]  [draw opacity=0] (8.93,-4.29) -- (0,0) -- (8.93,4.29) -- cycle    ;

\draw    (21.5,82.29) -- (178.5,82.29) ;

\draw [shift={(18.5,82.29)}, rotate = 0] [fill={rgb, 255:red, 0; green, 0; blue, 0 }  ][line width=0.08]  [draw opacity=0] (8.93,-4.29) -- (0,0) -- (8.93,4.29) -- cycle    ;
\draw    (22.5,100.29) -- (179.5,100.29) ;

\draw [shift={(19.5,100.29)}, rotate = 0] [fill={rgb, 255:red, 0; green, 0; blue, 0 }  ][line width=0.08]  [draw opacity=0] (8.93,-4.29) -- (0,0) -- (8.93,4.29) -- cycle    ;
\draw    (20.5,118.29) -- (177.5,118.29) ;
\draw [shift={(180.5,118.29)}, rotate = 180] [fill={rgb, 255:red, 0; green, 0; blue, 0 }  ][line width=0.08]  [draw opacity=0] (8.93,-4.29) -- (0,0) -- (8.93,4.29) -- cycle    ;

\draw    (23.5,136.29) -- (180.5,136.29) ;

\draw [shift={(20.5,136.29)}, rotate = 0] [fill={rgb, 255:red, 0; green, 0; blue, 0 }  ][line width=0.08]  [draw opacity=0] (8.93,-4.29) -- (0,0) -- (8.93,4.29) -- cycle    ;
\draw    (19.5,154.29) -- (176.5,154.29) ;
\draw [shift={(179.5,154.29)}, rotate = 180] [fill={rgb, 255:red, 0; green, 0; blue, 0 }  ][line width=0.08]  [draw opacity=0] (8.93,-4.29) -- (0,0) -- (8.93,4.29) -- cycle    ;

\draw    (204.5,19.29) -- (204.53,260.67) ;

\draw    (368.5,19) -- (367.87,258.67) ;

\draw    (206.5,31.29) -- (363.5,31.29) ;
\draw [shift={(366.5,31.29)}, rotate = 180] [fill={rgb, 255:red, 0; green, 0; blue, 0 }  ][line width=0.08]  [draw opacity=0] (8.93,-4.29) -- (0,0) -- (8.93,4.29) -- cycle    ;

\draw    (208.5,49.29) -- (365.5,49.29) ;

\draw [shift={(205.5,49.29)}, rotate = 0] [fill={rgb, 255:red, 0; green, 0; blue, 0 }  ][line width=0.08]  [draw opacity=0] (8.93,-4.29) -- (0,0) -- (8.93,4.29) -- cycle    ;
\draw    (208.5,82.29) -- (365.5,82.29) ;

\draw [shift={(205.5,82.29)}, rotate = 0] [fill={rgb, 255:red, 0; green, 0; blue, 0 }  ][line width=0.08]  [draw opacity=0] (8.93,-4.29) -- (0,0) -- (8.93,4.29) -- cycle    ;
\draw    (205.5,67.29) -- (362.5,67.29) ;
\draw [shift={(365.5,67.29)}, rotate = 180] [fill={rgb, 255:red, 0; green, 0; blue, 0 }  ][line width=0.08]  [draw opacity=0] (8.93,-4.29) -- (0,0) -- (8.93,4.29) -- cycle    ;

\draw    (18.5,170.29) -- (175.5,170.29) ;
\draw [shift={(178.5,170.29)}, rotate = 180] [fill={rgb, 255:red, 0; green, 0; blue, 0 }  ][line width=0.08]  [draw opacity=0] (8.93,-4.29) -- (0,0) -- (8.93,4.29) -- cycle    ;

\draw    (179.5,178.29) -- (20.5,179.27) ;
\draw [shift={(17.5,179.29)}, rotate = 359.65] [fill={rgb, 255:red, 0; green, 0; blue, 0 }  ][line width=0.08]  [draw opacity=0] (8.93,-4.29) -- (0,0) -- (8.93,4.29) -- cycle    ;

\draw    (18.5,186.29) -- (175.5,186.29) ;
\draw [shift={(178.5,186.29)}, rotate = 180] [fill={rgb, 255:red, 0; green, 0; blue, 0 }  ][line width=0.08]  [draw opacity=0] (8.93,-4.29) -- (0,0) -- (8.93,4.29) -- cycle    ;

\draw    (19.5,202.29) -- (176.5,202.29) ;
\draw [shift={(179.5,202.29)}, rotate = 180] [fill={rgb, 255:red, 0; green, 0; blue, 0 }  ][line width=0.08]  [draw opacity=0] (8.93,-4.29) -- (0,0) -- (8.93,4.29) -- cycle    ;

\draw    (23.5,218.29) -- (180.5,218.29) ;

\draw [shift={(20.5,218.29)}, rotate = 0] [fill={rgb, 255:red, 0; green, 0; blue, 0 }  ][line width=0.08]  [draw opacity=0] (8.93,-4.29) -- (0,0) -- (8.93,4.29) -- cycle    ;
\draw  [dash pattern={on 4.5pt off 4.5pt}]  (19.5,34.29) -- (176.5,34.29) ;
\draw [shift={(179.5,34.29)}, rotate = 180] [fill={rgb, 255:red, 0; green, 0; blue, 0 }  ][line width=0.08]  [draw opacity=0] (8.93,-4.29) -- (0,0) -- (8.93,4.29) -- cycle    ;

\draw  [dash pattern={on 4.5pt off 4.5pt}]  (180.5,42.29) -- (21.5,43.27) ;
\draw [shift={(18.5,43.29)}, rotate = 359.65] [fill={rgb, 255:red, 0; green, 0; blue, 0 }  ][line width=0.08]  [draw opacity=0] (8.93,-4.29) -- (0,0) -- (8.93,4.29) -- cycle    ;

\draw  [dash pattern={on 4.5pt off 4.5pt}]  (19.5,50.29) -- (176.5,50.29) ;
\draw [shift={(179.5,50.29)}, rotate = 180] [fill={rgb, 255:red, 0; green, 0; blue, 0 }  ][line width=0.08]  [draw opacity=0] (8.93,-4.29) -- (0,0) -- (8.93,4.29) -- cycle    ;

\draw    (393.5,19.29) -- (393.87,258.67) ;

\draw    (557.5,19) -- (555.53,256) ;

\draw    (395.5,31.29) -- (552.5,31.29) ;
\draw [shift={(555.5,31.29)}, rotate = 180] [fill={rgb, 255:red, 0; green, 0; blue, 0 }  ][line width=0.08]  [draw opacity=0] (8.93,-4.29) -- (0,0) -- (8.93,4.29) -- cycle    ;

\draw    (397.5,49.29) -- (554.5,49.29) ;

\draw [shift={(394.5,49.29)}, rotate = 0] [fill={rgb, 255:red, 0; green, 0; blue, 0 }  ][line width=0.08]  [draw opacity=0] (8.93,-4.29) -- (0,0) -- (8.93,4.29) -- cycle    ;

\draw (99,56) node  [font=\footnotesize] [align=left] {{\footnotesize Client Hello + Params}};
\draw (99,73) node  [font=\footnotesize] [align=left] {{\footnotesize Server Hello + Params + Cert}};
\draw (100,91) node  [font=\footnotesize] [align=left] {{\footnotesize Request for Client Cert}};
\draw (101,109) node  [font=\footnotesize] [align=left] {{\footnotesize Client Cert + Params + Session Key}};
\draw (101,127) node  [font=\footnotesize] [align=left] {{\footnotesize Server Confirmation}};
\draw (100,145) node  [font=\footnotesize] [align=left] {{\footnotesize Client Confirmation}};
\draw (101,163) node  [font=\footnotesize] [align=left] {{\footnotesize TCP 3-Way Handshake}};
\draw (102,195) node  [font=\footnotesize] [align=left] {{\footnotesize Req.}};
\draw (103,211) node  [font=\footnotesize] [align=left] {{\footnotesize Resp.}};
\draw (97,26) node  [font=\footnotesize] [align=left] {{\footnotesize TCP 3-Way Handshake (optional)}};
\draw (284,40) node  [font=\footnotesize] [align=left] {{\footnotesize Sever Rejection + Cert}};
\draw (286,75) node  [font=\footnotesize] [align=left] {{\footnotesize Server Hello + Resp.}};
\draw (283,24) node  [font=\footnotesize] [align=left] {{\footnotesize Client Hello (Invalid) + Cert + Req.}};
\draw (283,59) node  [font=\footnotesize] [align=left] {{\footnotesize Client Hello (Valid) + Cert + Req.}};
\draw (472,24) node  [font=\footnotesize] [align=left] {{\footnotesize Client Hello (Valid) + Cert + Req.}};
\draw (473,40) node  [font=\footnotesize] [align=left] {{\footnotesize Server Hello + Resp.}};
\draw (99,253) node   [align=left] {{\small \textit{Notional TLS-based VPN}}};
\draw (286,251) node   [align=left] {{\small \textit{QUIC to Unknown Server}}};
\draw (475,252) node   [align=left] {{\small \textit{QUIC to Known Server}}};

\end{tikzpicture}
	\caption{Simplified Comparison of QUIC and VPN Initialization}
		\label{fig:quic_vs_vpn}
\end{figure*}

First, the initial QUIC connection can be negotiated in a single round-trip, substantially shorter than the TCP three-way handshake. When compared with alternative encrypted tunnel schemes - such as TLS-based VPNs - QUIC offers a substantial reduction in round-trip transfers (see Figure~\ref{fig:quic_vs_vpn}). Indeed, for previously known QUIC servers, it is possible for a client to begin transmitting data from the very first packet, enabling zero round-trip session initialization.

Additionally, unlike TCP, QUIC does not require that all packets in a stream be processed in a particular order - removing head-of-line blocking issues involved when losing packets in a TCP stream and permitting heavy multiplexing. This makes it possible for QPEP to encapsulate multiple TCP flows inside a single QUIC session. This further reduces the number of session-initialization round-trips required to handle concurrent TCP streams. 

Like TCP, QUIC has built-in support for the re-transmission of lost and corrupted packets. This obviates many of the historical barriers to using UDP-tunnels over the satellite link as QUIC provides guaranteed packet transmission. Moreover, some draft proposals suggest the addition of built in forward error correction (FEC). These proposals have largely stalled as initial experiments found only minimal performance differences in terrestrial networks from FEC~\cite{swettQUICFECV12016}. However, as the standard evolves and future work is done on QUIC FEC, QPEP may be able to take advantage of this feature for further reliability gains in the lossy satellite environment.

\subsubsection{Satellite Performance}
As QUIC is a relatively recent protocol, its use in satellite environments has not been subject to much research. What little does exist is largely inconclusive. Some preliminary real-world assessments have found significant issues with QUIC performance in satellite environments resulting in an up to 100\% increase in page load times compared to PEP-accelerated TCP connections~\cite{thomasGoogleQUICPerformance2019}. However, other assessments suggest that QUIC performs far better than TCP, measuring up to a 50\% decrease page load times~\cite{rulaMileHighWifi2018}. Preliminary IETF and industry discussions have lead to a number of proposed (but unimplemented) techniques for optimizing QUIC connections to perform over satellite links~\cite{kuhnQUICSATCOM2019}.

This research on QUIC performance in satellite environments has focused on real-world end-to-end QUIC connections used to access HTTP2 web-servers which support the QUIC protocol. In these cases, researchers only control the client-side QUIC configuration settings. However, as QPEP operates in a distributed model, it may be possible to modify QUIC configuration settings or even implementation details to optimize both server and client parameters for the satellite link. Much as many modern PEP products modify TCP congestion windows on the satellite link, QPEP may tailor QUIC parameters relating to FEC, ACK decimation and congestion control to optimize for the satellite environment.

\subsection{QPEP Implementation}
Beyond the QUIC tunnel, QPEP operates in a manner similar to a standard unencrypted distributed PEP. A typical QPEP installation consists of two applications: the QPEP client which is located on the customer side of the satellite link and the QPEP server which is located on the internet side of the satellite link - either at the satellite gateway (as in the traditional PEP model) or elsewhere on the web.

The QPEP client may be installed anywhere on the network path prior to the satellite hop. Typically, this would likely be at the customer's local router. At startup, the QPEP client negotiates a long-standing QUIC session with the QPEP server - typically requiring one round-trip for key negotiation. After this, the QPEP client intercepts and terminates all incoming TCP connections, ``spoofing'' the destination TCP server and completing both the three-way handshake and all ACK messaging operations locally. A new QUIC stream is opened and mapped to each TCP session. As these streams are multiplexed within the overall QUIC session, no additional handshake is required over the satellite link.

Within the QUIC stream, TCP payloads and relevant header information are encapsulated into QUIC packets and transmitted across the satellite hop. For simplicity, only IP address and port fields are maintained, but there are no technical barriers preventing the transmission of other TCP header data (e.g. checksums or control flags). A socket listener on the QPEP server parses this header information and payload as it arrives. If no existing TCP session managed by the QPEP server matches the TCP four-tuple extracted from the packet header (i.e. Source Port, Source IP, Destination Port, and Destination IP), the QPEP server initiates a new TCP session based on this information. Otherwise, the TCP payload is routed to the appropriate existing TCP session.

When a response is received from the internet, this same process occurs in reverse. The QPEP server transmits an ACK immediately and then encapsulates relevant TCP headers and payload data into the active QUIC stream associated with the TCP conversation. If no existing TCP session matches the extracted headers (e.g. this is an attempt to initialize a session from the internet to the satellite client), a new QUIC stream is opened with the QPEP client and the three-way handshake is completed immediately on the internet-side of the satellite link. The QPEP client then decapsulates received packets, converts them back into TCP packets and routes them across the satellite customer's local network.

QPEP is designed as a transparent PEP - requiring no special software or configuration from connected hosts. This differentiates it from application-layer commercial PEPs which have limited compatibility with IOT and embedded systems. While this proof-of-concept implementation of QPEP focuses only on TCP connections targeted by traditional PEPs, only minor modifications would be required to add support for other protocols. This may be desirable for customers seeking to encrypt UDP or ICMP messages over the satellite link.

An open-source reference implementation of the QPEP approach is available in conjunction with this paper. This implementation is written using Go. Go was selected to increase accessibility without substantial performance sacrifices and to facilitate future work on these problems. To the best of our knowledge, only two non-proprietary PEP implementations exist, both of which are implemented in C/C++, have no support for encrypted communications, and have received only minimal development attention over the past several years~\cite{lacameraPEPsalTCPPerformance2016,delannoyTCPeP2013}. Source artifacts for other academic PEPs are either not publicly available or are restricted to particular simulation environments (e.g. NS-2)~\cite{duquerroySatiPSecOptimizedSolution2004, velenisSaTPEPTCPPerformance2002}. 

The QUIC implementation used by QPEP is based on the widely used quic-go library which roughly tracks the IETF proposal for QUIC standards~\cite{clementeQuicgo2019}. As discussed in Section~\ref{sec:quic-optimizations}, minor optional modifications to the QUIC implementation are made to optimize performance in the satellite networking environment. Future work considering the suitability of other QUIC implementations - such as Google's Chromium implementation - may prove valuable~\cite{googleQUICMultiplexedStream2020}.

\section{Secure PEP Testbed}
One of the principal challenges in developing and evaluating PEP appliances has been the difficulty in creating realistic simulations of satellite networking systems. The OpenSAND engine, previously Platine, is a long-standing satellite network simulation environment for replicating satellite data operations~\cite{viveristechnologiesOpenSAND2019}. OpenSAND supports modeling of both static latency scenarios that would be expected in GEO broadcasting as well as variable latency typical of LEO constellations. Moreover, while many prior simulations of TCP performance in high-delay networks simply rely on the addition of artificial delay to terrestrial networks, OpenSAND facilitates more realistic metrics. For example, the engine supports built in attenuation modeling and SNR emulation - facilitating realistic packet-loss conditions which have significant impacts on TCP congestion control performance. OpenSAND emulates the satellite network down to the link layer, simulating low-level protocol mitigations to noise and packet loss and creating more real-world conditions.

OpenSAND, while a powerful simulation tool, is somewhat difficult to configure - requiring multiple devices and precise network conditions to deploy in its standard setup - a barrier which has been noted in prior work~\cite{augerMakingTrustableSatellite2019}. In the process of developing our own simulations, we have created a dockerized deployment of the OpenSAND engine geared towards consistent and replicable evaluation of secure PEP implementations. This hopefully simplifies the process for researchers interested in making future related contributions. 

The testbed deployment creates a simple satellite network with a single gateway and satellite user terminal (as in Figure~\ref{fig:bent_pipe} and Figure~\ref{fig:qpep_overview}). The gateway is connected by default to the broader internet which allows a user connected to the satellite terminal to open a web-browser and visit real-world web services over the simulated satellite link. A gateway workstation is also provided to facilitate reliable performance metrics on the emulated host network with common benchmarking tools pre-installed. Additionally, packet-capture tools on the satellite node enable real-time monitoring of over-the air transmissions, allowing for verification that sensitive data is not transmitted in clear-text. Pre-configured installations of QPEP, OpenVPN, and PEPsal are provided on both the gateway and satellite terminal networks. A set of example python scripts are provided to launch various benchmarking tools and orchestrate the simulation scenarios used in our own performance evaluations. These scripts are designed as modular components to facilitate the direct and consistent comparison of QPEP with proposals made in future work.

\section{Q-PEP Simulated Evaluation}
In this section, we present an evaluation of the QPEP approach to providing secure over-the-air satellite broadband transmissions and its impact on the performance of TCP-based traffic. To the best of our knowledge, no comparable encrypted satellite PEP is publicly available for evaluation. As such, we selected PEPsal, an open-source unencrypted PEP solution, and OpenVPN, an open-source VPN product without specific satellite networking optimizations, to provide some context to measurements made. First, we present simple experimental results under the default OpenSAND network conditions. Next, we consider how QPEP fares under varying satellite network conditions - namely high rates of packet loss and in LEO networks with variable delay. Finally, we evaluate the performance impacts of various modifications to the QUIC protocol used in the PEP tunnel. All simulations are executed on the same laptop (Intel Core i7-7770HQ @2.8GHz, 32GB RAM, Windows 10) and the testbed environment is re-initialized between scenario runs.

Unless otherwise noted, the OpenSAND network is configured to use the DVB-S2 protocol with GSE encapsulation for forward-link communications and DVB-RCS2 with RLE encapsulation for the return link. The clear-sky SNR is set to 20~dB and Adaptive Coding and Modulation (ACM) is used at the physical layer to provide quasi error free (QEF) communications at this SNR level. A constant speed-of-light delay of 125~ms is used from both the satellite terminal and the satellite gateway to the satellite (resulting in a 500~ms RTT). The forward-link carrier frequency is allocated 50.0~MHz of bandwidth with a roll-off factor of 0.25 and the return-link is allocated approximately 7.4~MHz of bandwidth.

Simulations of QPEP are configured with a QPEP server sitting local to the satellite gateway network and listening for incoming QUIC tunnel connections. The QPEP client is hosted on the satellite terminal and listens transparently into all incoming TCP connections. The QPEP server is configured to accept up to 40,000 concurrent streams from a single host - substantially higher than the go-quic library's default setting of 100. This is to enable compatibility with concurrent benchmarking downloads.

OpenVPN simulations are deployed similarly to QPEP, with an OpenVPN client connected to the satellite terminal and an OpenVPN server connected to the satellite gateway.

PEPsal is evaluated under two different configurations - a distributed installation and an integrated installation. Evaluations of distributed PEPsal are implemented with a PEPsal endpoint transparently listening to all incoming TCP traffic on both the satellite gateway and the satellite terminal. In integrated PEPsal, a PEPsal endpoint listens to incoming TCP traffic on the satellite terminal but no endpoint is installed on the satellite gateway. The effect of integrated PEPsal could thus be realized without ISP participation while the distributed installation would require explicit ISP support.

\subsection{Basic Performance Evaluations}
\label{sec:basic_evals}
\begin{figure}
	\includegraphics[width=\linewidth]{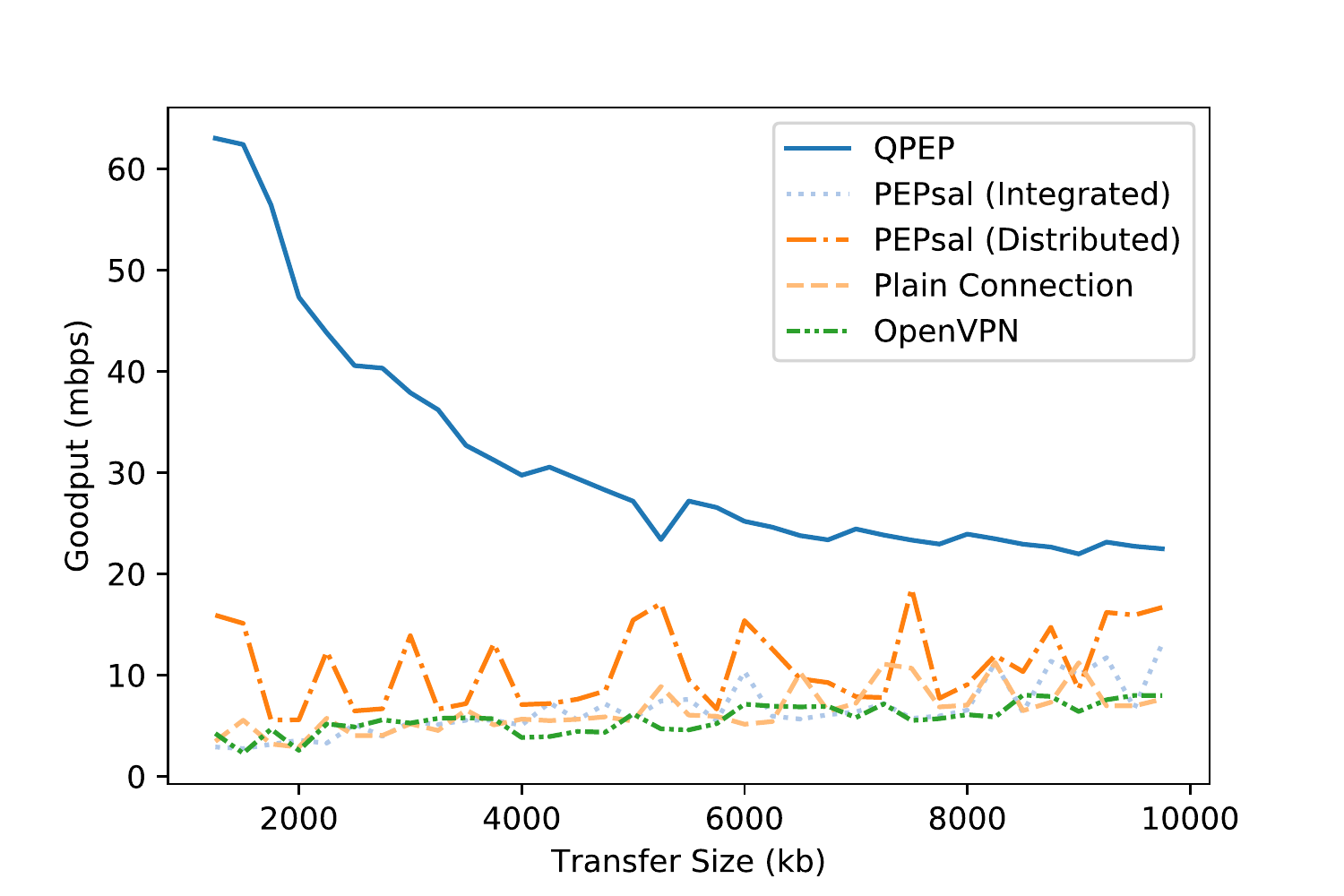}
	\caption{Goodput Comparison by Iperf Transfer Size}
	\label{fig:iperf_basic}
\end{figure}

An initial comparative assessment of goodput can be made through the use of the Iperf benchmarking tool which attempts to provide consistent performance evaluations of network speed. For these benchmarks, an Iperf server is hosted on the satellite gateway network and is used to transfer data to an Iperf client connected to the satellite terminal network. Several iterations of Iperf are run, with data transfer sizes ranging from 0.5 to 10~mb. Varying the volume of data transferred provides important insights into the extent to which results are influenced by session initialization time or TCP slow-start dynamics that exert heavy influence on small file sizes but which matter less as size increases. The results of these simulations are summarized in Figure~\ref{fig:iperf_basic}.

These results suggest that QPEP is capable of making significantly greater use of bandwidth for moderately-sized file downloads under the Iperf benchmark than any of the tested alternatives. To some extent, this is expected as Iperf generates many concurrent TCP connections as part of its test. As QPEP takes advantage of QUIC's internal support for multiplexing to bundle these concurrent TCP streams into a single QUIC session, it is unsurprising that measured goodput is significantly higher than in approaches which lack multiplexing. Additionally, as QUIC supports sending data along with the session initialization information in the first packets, QPEP performs well for the transfer of very small files which can be completed often in the first round-trip. 

Integrated PEPsal does not offer significant advantage in this use-case as it is still constrained by head-of-line blocking over the satellite hop and a large portion of traffic is on the un-optimized route from the gateway to the user terminal. Distributed PEPsal performs better as it is able to optimize both directions of the satellite conversation. However, it also does not match QPEP's ability to encapsulate concurrent downloads. Finally, as expected, OpenVPN is the worst performing of all options with no particular optimizations for the satellite link and additional overhead from the VPN layer making it perform slightly worse than a plain satellite connection.

This benchmark, while meaningful, is somewhat misleading. Iperf provides one important measure of goodput but the scenario it evaluates is not representative of real-world satellite internet user behavior. Specifically, opening a connection to a port, ramping it up to maximum speed, and then maintaining that speed for many file transfers is not how most web services operate. Very few individuals are, for example, streaming video or downloading large amounts of data over expensive and low-bandwidth satellite links. PEPs were invented to optimize web-browsing and visits to mostly text and image-based online services. Even if QPEP were a useful tool for encrypting large file transfers over satellite, its adoptability in practice hinges on the ability to match existing PEPs for typical use cases.

\begin{figure}[h]
	\includegraphics[width=\linewidth]{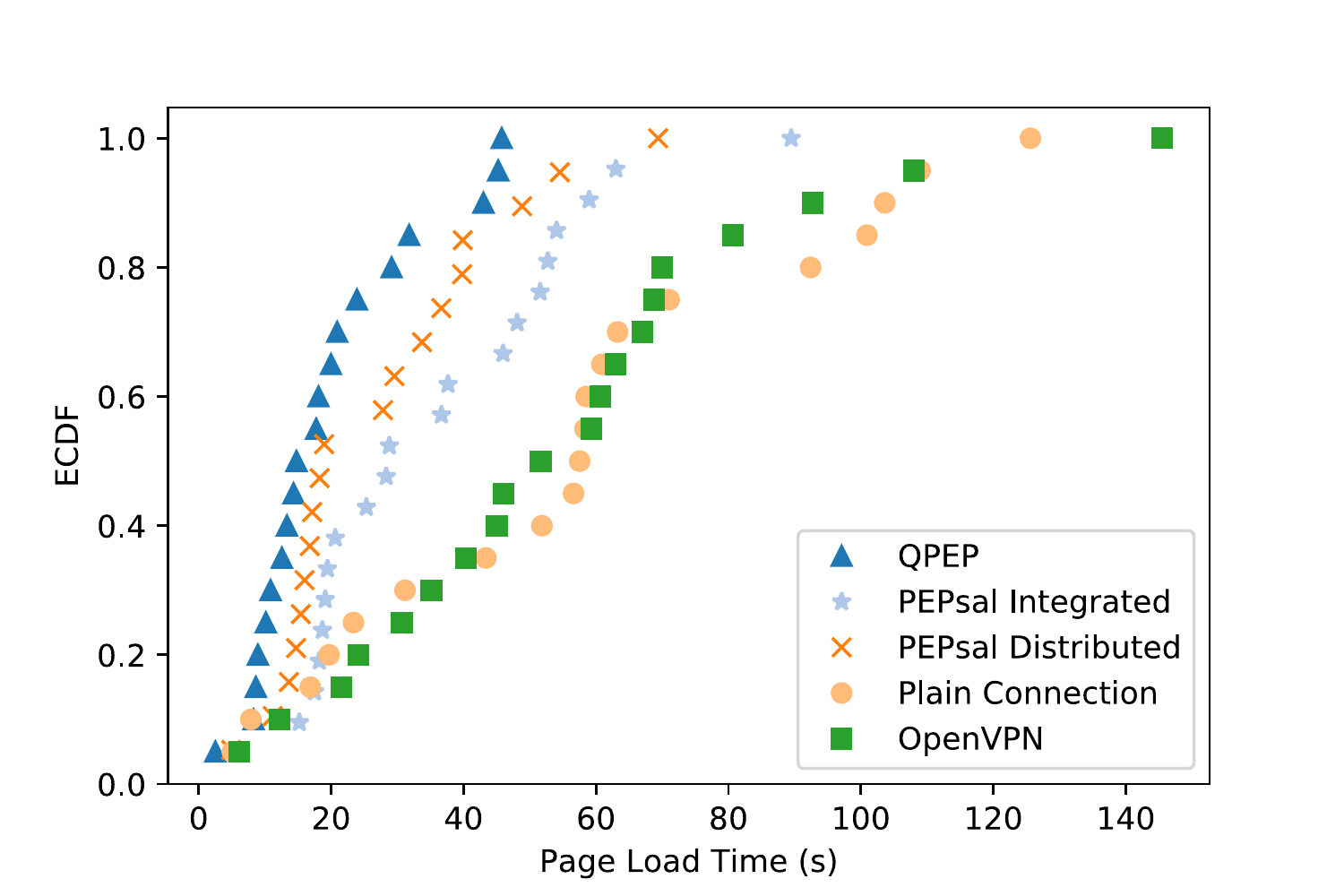}
	\caption{ECDF Comparison of PLTs over Alexa Top 20}
	\label{fig:plt_basic}
\end{figure}

A more realistic sense of this ability can be found through the evaluation of the time it takes to visit real-world websites. Unlike the IPerf benchmark, web-browsing consists of the transfer of many small files (e.g. embedded images or style-sheets) over multiple TCP sessions. Often, these files can be hosted on a variety of servers. This makes web traffic more sensitive to satellite latency effects than simple file transfer traffic.

Experimentally measuring page load times is an imprecise art. For our simulations, we used the open-source web benchmarking tool Browsertime~\cite{sitespeedioBrowsertime2019}. Browsertime reports page load time as the number of elapsed miliseconds from the initial web browser request to the navigationStart event of the browser's underlying navigation timing API~\cite{mozilla.orgPerformanceTimingNavigationStart219}. In our simulations, the satellite gateway was connected to a terrestrial broadband network so as to connect to real-world websites. This creates some variability between measurement runs as they depend heavily on network conditions at measurement time. To reduce this variability, measurements were made of connections to the top 20 distinct domains listed by Alexa Internet Inc~\cite{alexaTop500SitesWeb2020}. Between each website visit, the browser cache was reset. The results of these page load time measurements are summarized by means of an Empirical Cumulative Distribution Function (ECDF) in Figure~\ref{fig:plt_basic}.

This page load time comparison shows that QPEP provides encrypted TCP sessions under real-world use-scenarios without harming performance. While QPEP's advantage here is less dramatic than in the IPerf benchmark, it still outperforms the unencrypted PEPsal proxy. QPEP's mean page load time (PLT) of 19.9~seconds is roughly 30\% faster than distributed PEPsal's 27.8~seconds and almost 45\% faster than integrated PEPsal's 35.9~second mean PLT. 

The disadvantage of using traditional VPN technology to encrypt satellite communications is quite clear as well. OpenVPN takes, on average, almost thrice as long to load web pages compared to QPEP, with a mean PLT of 56.4~seconds. This suggests that QPEP meets its primary purpose of providing an encrypted over-the-air alternative to traditional VPNs without incurring the same magnitude of performance loss. QPEP is significantly more performant in this benchmark than even unencrypted PEPs while offering higher degree of security and not requiring installation of any software on the ISP gateway itself.

\subsection{Satellite Network Variations}
While these basic evaluations present a compelling case for the use of the QPEP approach to provide communications security in typical GEO networks, many satellite networks demonstrate atypical characteristics. Packet loss, rain-fade, and variable routing delays can all have significant impacts on satellite network performance. As such, we have elected to evaluate the relative performance of QPEP under some of these conditions.

Intuitively, packet loss and rain fade conditions are likely significant challenges for the performance of tunneling PEPs like QPEP. Loss of critical packets related to they key exchange process or initialization of QUIC sessions could impose significant additional round-trip costs not observed in our clear-sky condition tests. Tunneling PEPs are also vulnerable to high-levels of packet loss which, depending on implementation, can cause the tunnel between the PEP client and server to timeout or otherwise break altogether. However, the use of a PEP can also improve network performance under loss conditions as it mitigates the impact of TCP congestion-control restarts as discussed in Section~\ref{sec:tcp_satellite_performance}.

Given these concerns, a series of simulations were run to assess QPEP's performance characteristics at low Signal to Noise Ratios (SNR). For these simulations, SNR was calculated by OpenSAND as a function of ``attenuation'' - a single value representing losses from various physical sources. The sum of attenuation values across each hop of the satellite link was subtracted from a ``clear-sky'' SNR (set at 20~dB). SNR values were tested at intervals of 0.5~dB, with 0.25~dB attenuation on the forward link to the satellite terminal and 0.25~dB attenuation on the forward link to the satellite ground station. To replicate real-world conditions, ACM was used at the physical layer to optimize modulation and coding settings (MODCOD) and reduce transmission errors. The MODCODs employed were derived from those used in standard DVB-S2/DVB-RCS2 implementations. SNRs from 20~dB (clear-sky) to 10~dB were measured. Lower ratios were not evaluated as ACM's ability to provide QEF diminished causing a large proportion of routing failures regardless of protocol and increasing simulation time significantly (due to IPerf's long timeouts).

\begin{figure}[h]
	\includegraphics[width=\linewidth]{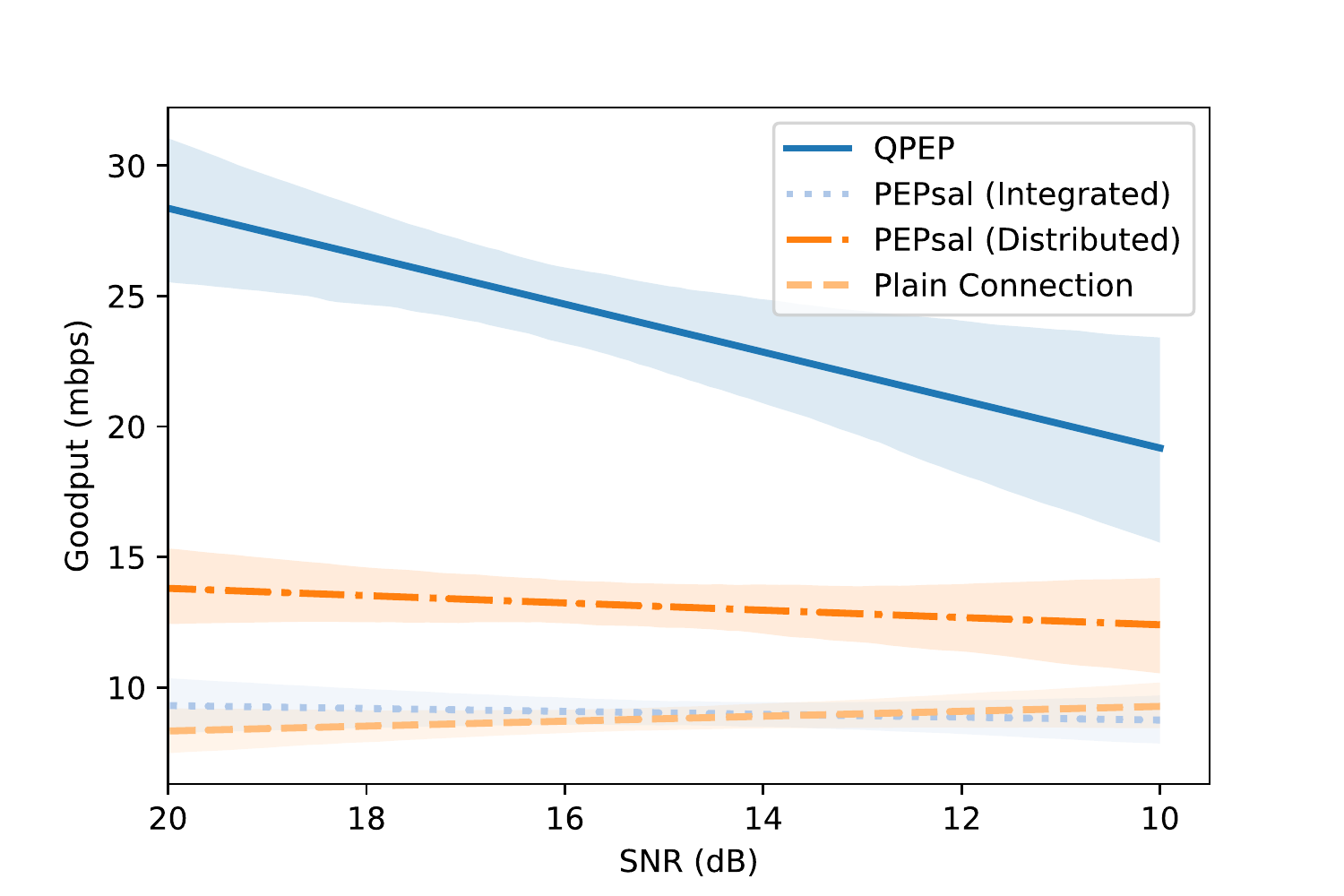}
	\caption{IPerf Performance at Sub-Optimal SNR}
	\label{fig:attenuation}
\end{figure}

At each SNR interval, performance was measured over five IPerf benchmarks with a file download size of 10~kb. Best fit linear regressions for the results of these approximately 400 simulation runs are presented in Figure~\ref{fig:attenuation} with shaded boundaries representing a 95\% confidence interval.

As expected, we find that QPEP performance suffers at diminishing SNR levels - albeit with increasing variability at very low SNRs. This makes sense as sessions which are ``lucky'' in the noisy environment and initialize without the loss of key handshake packets perform in-line with clear-sky rates while sessions which fail to do so face substantial RTT penalties. QPEP outperforms distributed PEPsal at all SNR levels but the inconsistency of its performance may make QPEP less suitable for situations where reliable bandwidth characteristics in the presence of noise are expected. Generally, the Iperf benchmark suggests that QPEP's  performance exceeds that of an unencrypted satellite connection, even at moderately low SNRs. At high levels of packet loss, however, this performance advantage diminishes significantly. While distributed PEPsal only failed to complete four IPerf sessions (one at 11~dB and three at 10.5~dB), QPEP dropped a total of seven (two at 10.5~dB and all five at 10~dB). The plain un-optimized connection dropped no sessions.

\begin{figure}
	\includegraphics[width=\linewidth]{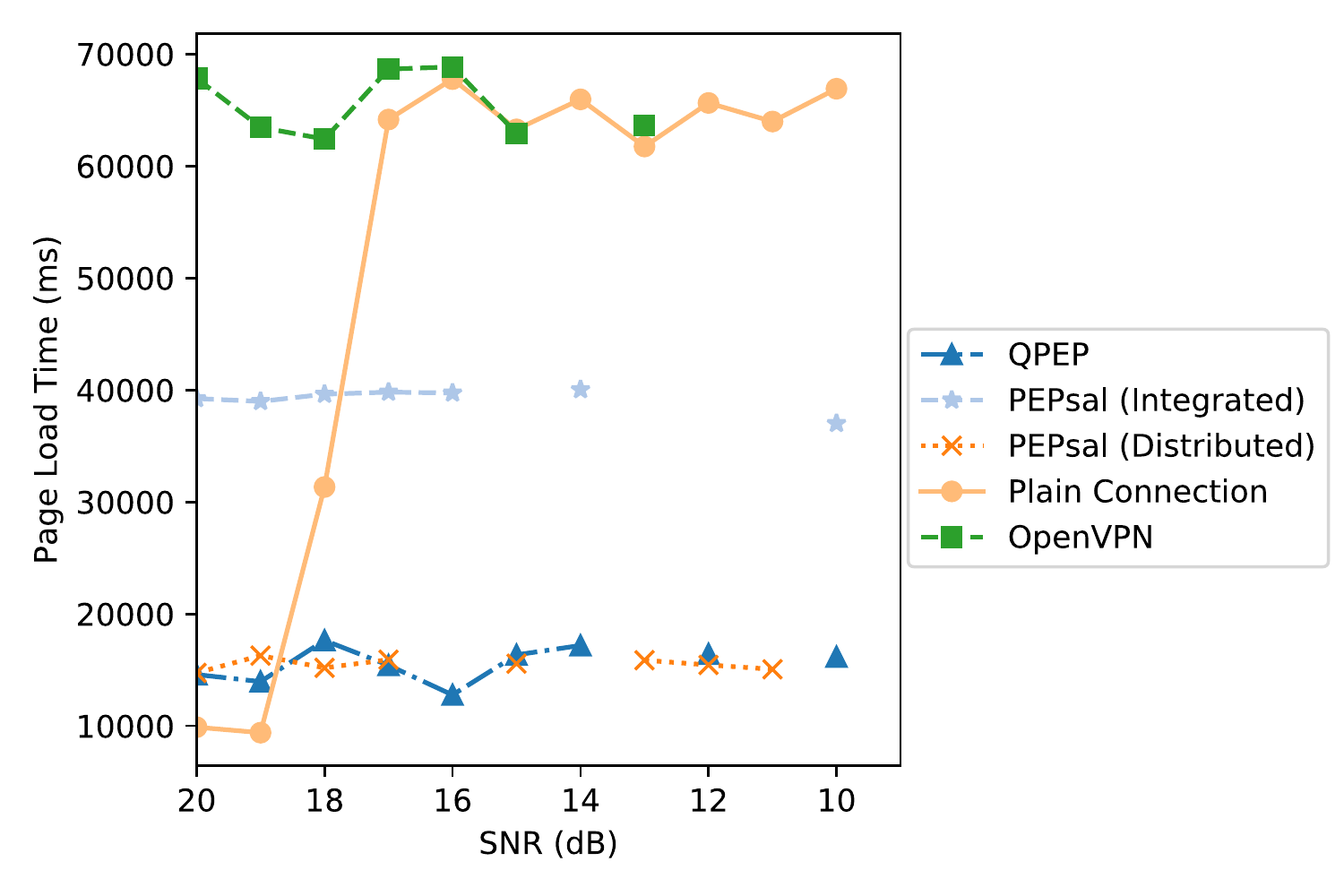}
	\caption{Average PLT of Amazon.com at Sub-Optimal SNR}
	\label{fig:attenuation_plt}
\end{figure}

Of course, while IPerf is a reasonable benchmark for file-transfer times, web-browsing performance is much more heavily influenced by the behavior of short small-data connections. These connections are likely more resilient to the effects of packet loss than the long-lived Iperf benchmark scenario. To assess the impact of attenuation on page load times, a series of simulations were run measuring the average page load time of the Amazon.com homepage over five visits at each SNR level (Figure~\ref{fig:attenuation_plt}). Amazon.com was selected as it is an e-commerce site with many embedded product images, resulting in numerous small and concurrent TCP-sessions of the sort PEPs are intended to optimize and is also one of the most performant sites we measured over an un-optimized satellite link in clear sky conditions. We find that this performance edge is quickly undermined in plain satellite networks as SNR (and TCP congestion control restarts) both increase. However, both QPEP and the distributed version of PEPsal maintain roughly equivalent performance at moderate SNR levels. QPEP avoided dropped connections until SNR fell below 14dB - outlasting distributed PEPsal which began to experience reliability issues at simulated SNRs of 17 and lower. 

At very low SNR levels, however, distributed PEPsal does appear to recover lost connections more reliably than QPEP. This is likely a result of corrupted QUIC session initialization packets (which are relatively large due to certificate data) preventing the initialization of an encrypted tunnel. As with the Iperf scenario, these results suggest that QPEP is suitable for mild to moderate SNR attenuation but may struggle in particularly harsh communications environments. It is possible a modified implementation of QPEP which uses preshared keys to enable zero RTT session initialization may be prove more resistant to this effect in future work.

\begin{figure}
	\includegraphics[width=\linewidth]{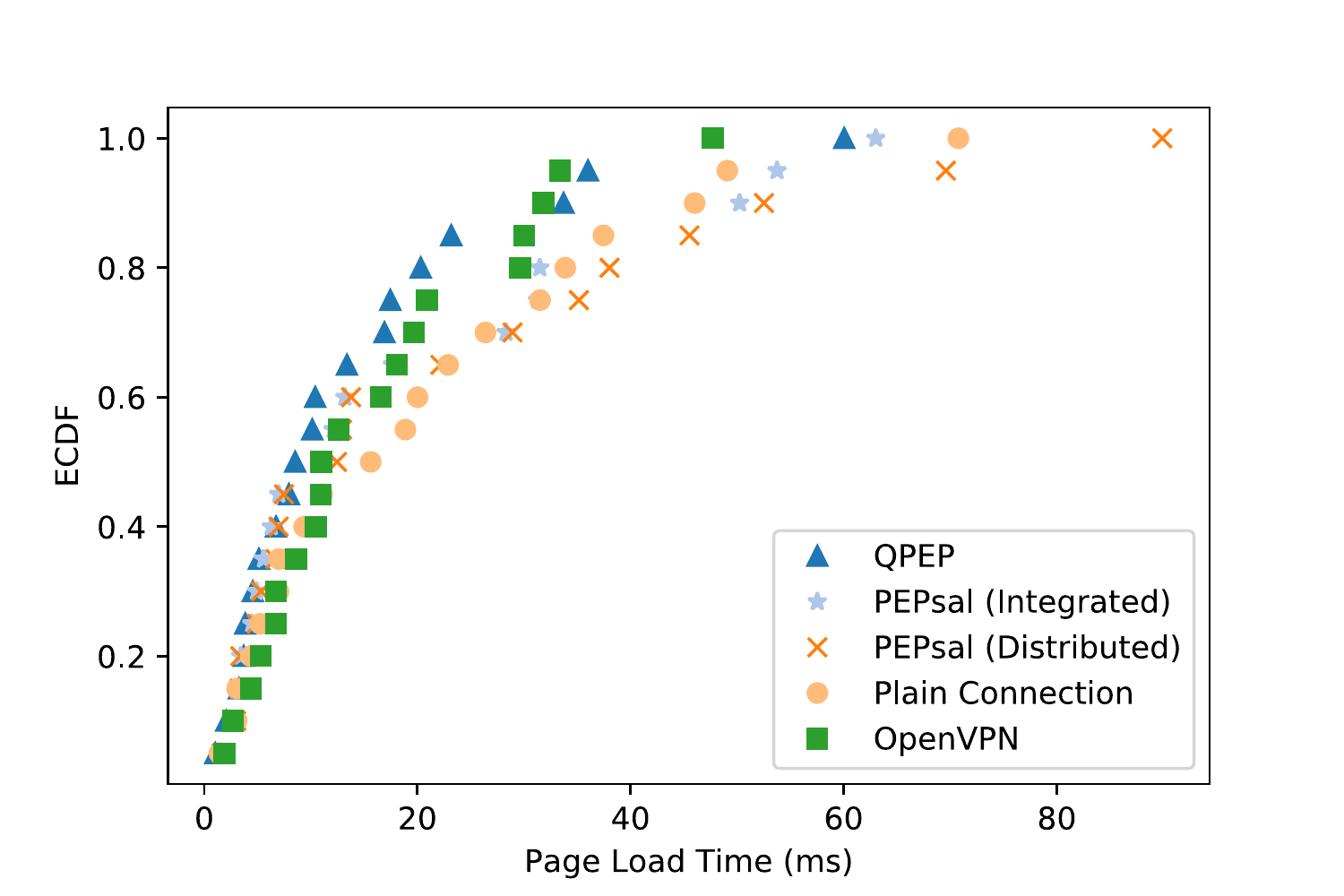}
	\caption{ECDF of Alexa Top 20 PLTs in Iridium Simulation}
	\label{fig:leo_plt}
\end{figure}

While this paper has focused on GEO satellite networks and performance under constant speed-of-light delays, many modern satellite broadband proposals focus on the use of LEO constellations which experience highly variable latency depending on the location and time of transmission. To assess performance under these conditions, we used an existing OpenSAND simulation model which replicates the delay characteristics of a satellite terminal in the Atlantic Ocean connecting through the Iridium LEO constellation to a gateway in London~\cite{boubakerOpenSANDExampleDelay2019}. In the emulated LEO network, one-way delay over the satellite hop varies from as low as 25~ms to as high as 140~ms depending on the time of transmission (and, by extension, the number of hops a signal must make through the Iridium constellation). The same page load time benchmark from section~\ref{sec:basic_evals} was run within the LEO network and the results are summarized in Figure~\ref{fig:leo_plt}.

As expected, the performance benefits of PEPs are less dramatic in LEO satellite networks and the disadvantage of VPN use significantly reduced. We find that QPEP still offers the best performance across these tests with a mean PLT of 14.4~seconds, besting OpenVPNs PLT of 16.4~seconds. OpenVPN may implement optimizations for the moderate levels of latency observed in this LEO scenario which lead to this improved performance. Meanwhile, PEPsal was roughly equivalent to an un-optimized connection with mean PLTs of 19.4~seconds (integrated) and 23.1~seconds (distributed) compared to a plain connection's 21.2~second mean PLT. It is possible that overhead incurred in PEPsal's implementation imposes performance costs in networks which lack substantial latency delays. Taken together, these measurements suggest that QPEP represents a viable mechanism for providing performant encryption in modern LEO constellations and possibly future mega-constellations. However, the performance gains over traditional VPNs in LEO are only marginal.

\subsection{QUIC Optimizations}
\label{sec:quic-optimizations}
One of the principal theoretical advantages of a distributed PEP configuration is the ability to adopt non-standard and environmentally tailored protocols over the satellite hop. In this section, we consider what modifications to the standard QUIC implementation may be well suited to the satellite environment and evaluate their impact on performance within our testbed. This is a cursory look and we expect that future work on optimizing the QUIC protocol for satellite - especially the congestion control mechanism - may prove fruitful.

One strategy for improving the performance of TCP ACKs over satellite links is ACK decimation - the process of combining many ACK messages into a single message sent at regular intervals. In the default QUIC implementation, ACK decimation is set to result in a minimum of one ACK message for every ten ACK-eliciting packets. However, as we control both ends of the QUIC link in a QPEP installation, it is possible to change this standard value on both the client and server side of the connection to optimize link behavior. By increasing number of elicitations per ACK, it may become possible to diminish the effect that link asymmetry has on the performance of the QPEP proxy. However, with greater spacing between ACK messages, the risk of unnoticed packet loss causing delay can increase.

These expectations are borne out by experimental simulations with QPEP using five runs of the IPerf benchmarking tool for a 10~mb file transfer at differing ACK decimation ratios ranging from 1 elicitation per ACK to 200. For the purpose of these simulations, ACK decimation began after the 10th packet over the QUIC link (as opposed to the default setting of 100 packets prior to ACK decimation) and timeout-based ACK queuing was disabled. These changes were made to better focus on the effect of ACK decimation in isolation without the complexities of the full QUIC congestion control implementation. As a result, measurements from these simulations are not directly comparable with those made in other sections. The IPerf benchmark was selected rather than PLT benchmarks as the longer-lived sessions would likely better exhibit the influence of high ACK decimation ratios than web-pages involving the transfer of relatively few packets. 

\begin{figure}[h]
	\includegraphics[width=\linewidth]{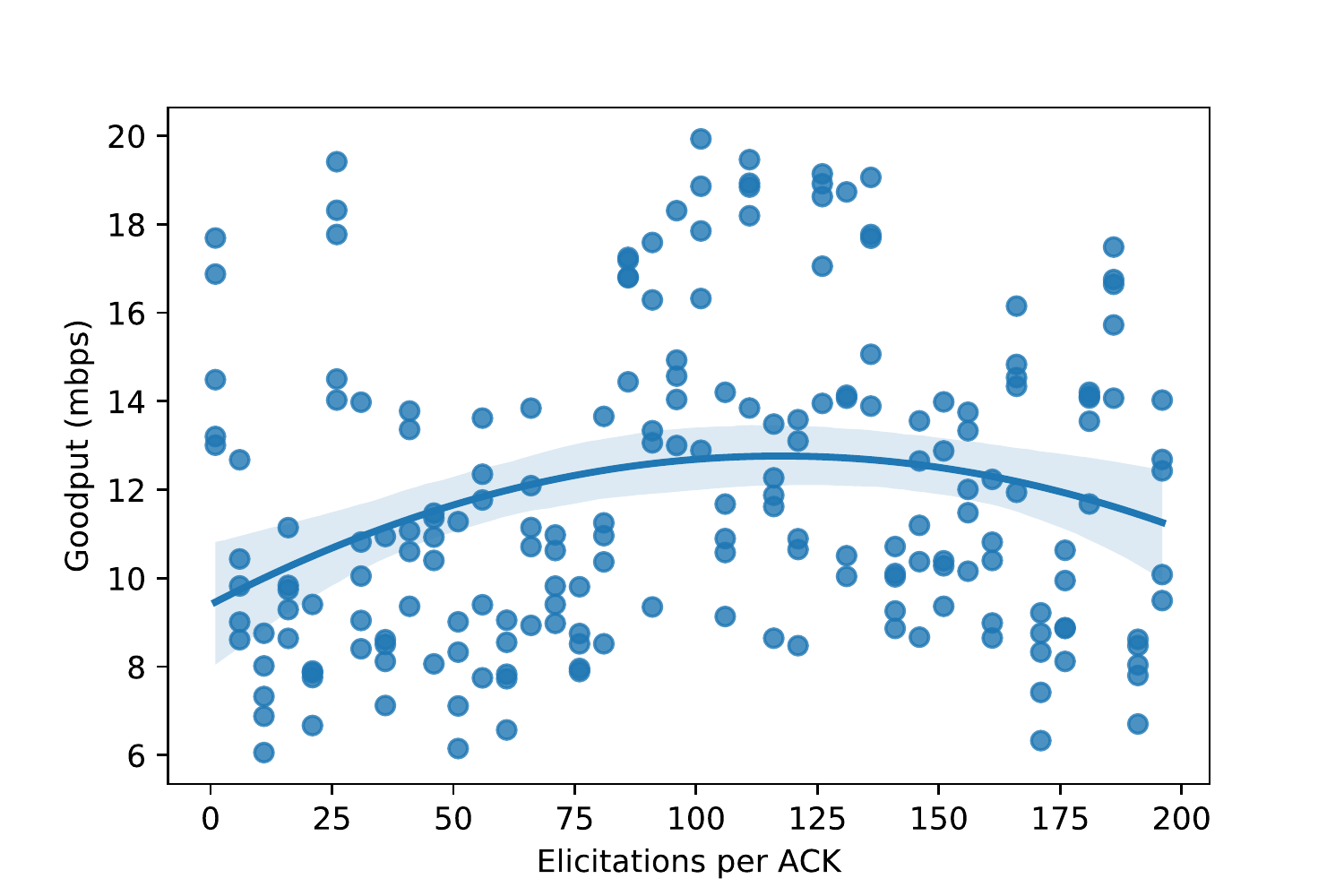}
	\caption{IPerf Goodput vs ACK Decimation Ratio}
	\label{fig:ack_decimation}
\end{figure}

The results of this simulation (see Figure~\ref{fig:ack_decimation}) suggest that greater ACK decimation ratios can result in increased goodput. However, at very high levels of ACK decimation, this relationship weakens as the risk of long-unnoticed packet loss leads to performance degradation. While these simulations were conducted under clear-sky dynamics, it is possible that optimal decimation ratios would vary substantially under adverse conditions. Future work geared towards analyzing this relationship to find an ideal (or dynamic) ACK decimation ratio would likely be productive - especially in a model which considers QUIC's other congestion control strategies. Still, this initial assessment shows that QPEP architecture permits bespoke QUIC optimizations not generally considered feasible in prior performance assessments of QUIC over satellite links. Even if QUIC were not, by default, sufficiently performant for satellite network encryption (e.g. in LEO constellations), modifications of this nature represent a promising mechanism for achieving further performance gains.

\begin{figure}[h]
	\includegraphics[width=\linewidth]{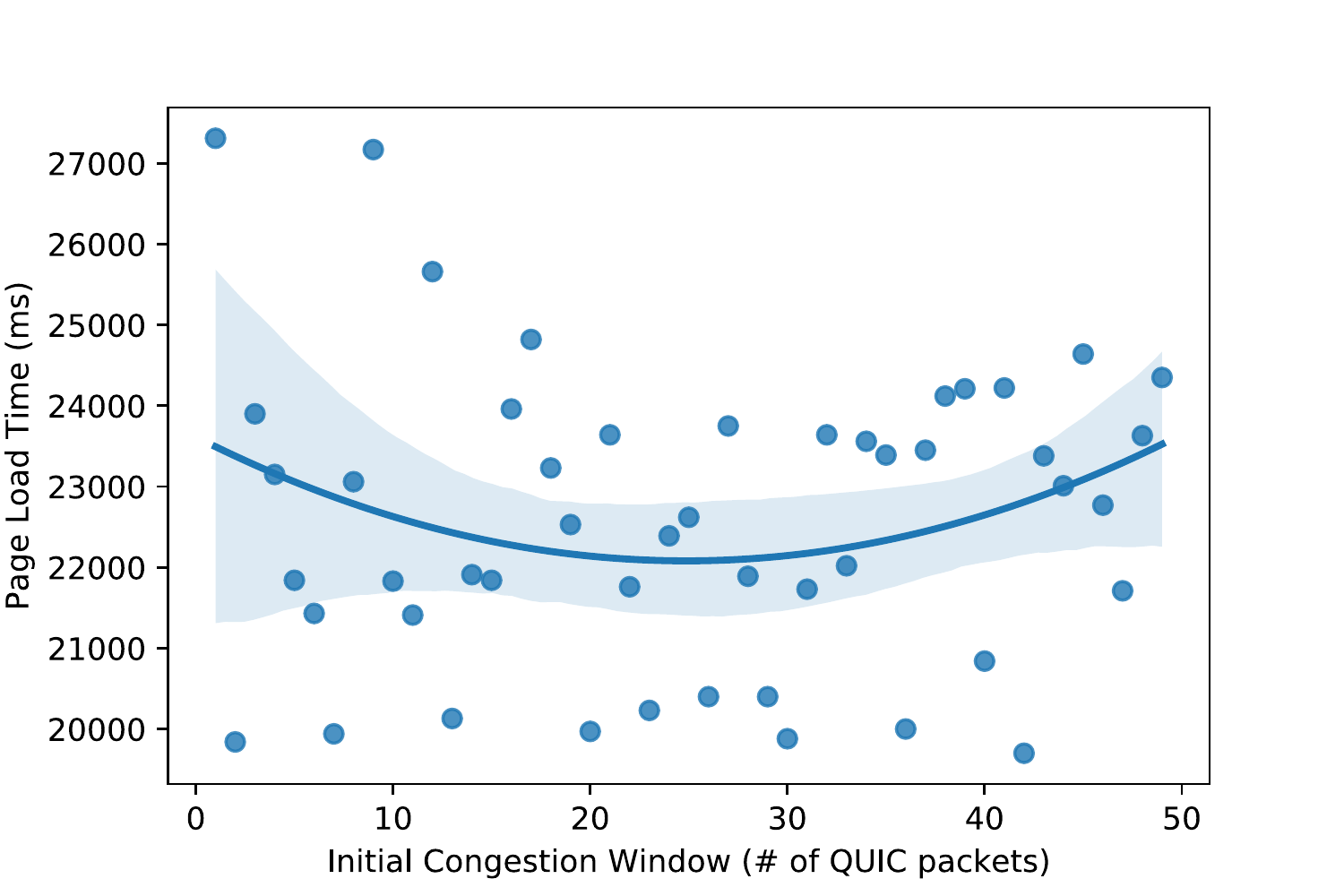}
	\caption{Average PLT of BBC Homepage vs Initial CWND size}
	\label{fig:cwnd_size}
\end{figure}

Another approach used in TCP PEPs is the adjustment of the initial congestion window (CWND) to accelerate the cumbersome TCP slow-start process. As QUIC implements a similar startup mechanism, it is possible that increasing the QUIC initial CWND size may have similar effects. In the standard QUIC implementation used for our simulations, this is set to the size of 10 QUIC packets. Several test visits to the BBC Homepage were run over initial CWND sizes ranging from 1 QUIC packet to 50. While polynomial regression suggests that there may exist an optimal initial CWND of around 25 QUIC packets, this association is not particularly pronounced (Figure~\ref{fig:cwnd_size}). At least for our PLT tests, QPEP appears to reach appropriate CWND size relatively quickly, regardless of initial starting condition.

These two demonstrative scenarios illustrate the case that modifications to the underlying QUIC protocol implementation used by QPEP may have a meaningful impact on proxy performance. There are dozens of constants and assumptions made in common QUIC protocol implementations and these are not necessarily tailored towards the unique network environment of satellite communications. Finding the perfect balance of QUIC parameters for satellite networking is well beyond the scope of this research and represents a significant but non-insurmountable engineering challenge. However, QPEP is already dramatically more performant than traditional VPN security over satellite and either equivalent to or better than unencrypted PEPs in our benchmarks. Thus, this optimization challenge, while interesting, is largely irrelevant to the overall suggestion that QUIC can contribute towards a secure and sufficiently performant TCP-PEP for modern satellite broadband.

\section{Future Work}

It is worth noting that the QPEP implementation presented here is very much a demonstrative proof-of-concept. Future work, either for academic purposes or for real-world commercial use would likely benefit from implementing several additional features. First, the networking protocols supported by QPEP are limited only to TCP over IPv4. Adding IPv6 support and support for tunneling non-TCP protocols (such as UDP) would be useful. It is unlikely that QPEP would significantly improve performance of protocols like UDP (as UDP performance is not as latency sensitive), but encrypting DNS requests and other information sent over UDP datagrams would be key to realizing the full security and privacy benefits of encrypted over-the-air SATCOMs. Additionally, further support for atypical TCP-header parameters and better handling of TCP error states in the case of packet loss may improve real-world customer experience.

Beyond these largely engineering-related areas for further development, several academic questions may be worth closer consideration. As we have briefly discussed in section~\ref{sec:quic-optimizations}, the QPEP architecture obviates many of the concerns raised in prior work regarding the suitability of QUIC to satellite environments by allowing for end-to-end optimization of QUIC protocol parameters. Many more advanced features of QUIC - such as use of 0-RTT session initialization or proposed FEC standards - were beyond the scope of this preliminary research. However, the testbed presented here may prove useful for assessing the implications of such changes under various satellite networking conditions. While our testbed was designed for the development of a more secure PEP platform, it may be well suited to the general problem of understanding and optimizing the interaction between QUIC and satellite networks.

\section{Conclusion}
In this research, we have challenged the historical assumption that security and performance must trade off in high-latency satellite networks. We have presented a new approach to encrypting TCP satellite communications over-the-air through the use of QPEP - a performance enhancing proxy which leverages the open QUIC protocol standard to provide an encrypted UDP tunnel for the satellite hop. By making minor bespoke modifications to the QUIC protocol, we have demonstrated that QUIC may be further optimized for the particularities of the satellite environment - obviating many of the performance concerns raised by prior work.

To evaluate QPEP, we have developed an experimental benchmarking suite built on the OpenSAND satellite simulation engine. This open-source research testbed facilitates replicable and meaningful comparison of TCP PEP performance over satellite networks. Our evaluations demonstrate that the QPEP approach offers comparable performance to unencrypted PEPs and dramatically outperforms VPN-based encryption to provide secure TCP communications over satellite links. 

To our knowledge, QPEP is the first open source PEP proof-of-concept with support for the encryption of arbitrary TCP traffic. Moreover, unlike many commercial secure PEPs, QPEP operation is fully independent to the satellite ISP, allowing individual satellite customers to run their own QPEP servers in the cloud and adopt the protocol without substantial changes to ISP infrastructure or the sharing of sensitive traffic metadata with the ISP. While QPEP is a rudimentary proof of concept and further development work would be required for reliability in critical commercial networks, these preliminary results suggest that the QUIC-tunneling PEP approach is a promising and straightforward alternative to proprietary satellite encryption standards.

As the next generation of satellite broadband projects begins to launch, there remains a critical need for ensuring the security and privacy of TCP communications in these networks without sacrificing performance. The QPEP proof-of-concept presented here demonstrates the ability of extending open and verifiable standards to meet this need. Beyond QPEP, the benchmarking and testbed approach demonstrated in our research makes a methodological contribution, hopefully paving the way for future replicable research towards the development of secure modern satellite broadband.
\section*{Availability}
\label{sec:artifacts}
Source code and documentation for both our QPEP implementation and our OpenSAND-based testbed environment are available publicly (https://github.com/pavja2/qpep). Example python scripts used to run all of the simulation scenarios presented in this paper are provided. 

\bibliographystyle{plain}
\bibliography{qpep}
\todos
\end{document}